\documentclass[12pt]{article}
\usepackage[centertags]{amsmath}
\usepackage{amsfonts}
\usepackage{amssymb}
\usepackage{amsthm}
\usepackage{newlfont}
\usepackage{stmaryrd}
\usepackage{mathrsfs}
\usepackage{palatino} 
\usepackage{fancyhdr}
\usepackage{graphicx}
\usepackage{fancybox}
\usepackage{multibox}
\usepackage{geometry}
\usepackage[dvips]{epsfig}
\usepackage{eepic}
\usepackage{pst-plot}

  \let\La=\Lambda

\newcommand{\bbZ}{{\mathbb Z}}

\newcommand{\opunit}{\text{1}\kern-0.22em\text{l}}

\newcommand{\bde}{\begin{definition}}
\newcommand{\ede}{\end{definition}}
\newcommand{\beq}{\begin{equation}}
\newcommand{\eeq}{\end{equation}}
\newcommand{\ben}{\begin{enumerate}}
\newcommand{\een}{\end{enumerate}}
\newcommand{\ble}{\begin{lemma}}
\newcommand{\ele}{\end{lemma}}
\newcommand{\bpr}{\begin{proof}}
\newcommand{\epr}{\end{proof}}

\title{Finite-size effects for anisotropic bootstrap percolation:\\ logarithmic corrections.}

\author{
  {\normalsize Aernout C.~D.~van Enter} \\[-1mm]
  {\normalsize\it Institute for Mathematics and Computing Sciences} \\[-1.5mm]
  {\normalsize\it Rijksuniversiteit Groningen} \\[-1.5mm]
  {\normalsize\it Blauwborgje 3} \\[-1.5mm]
  {\normalsize\it P.O.Box 800} \\[-1.5mm]
  {\normalsize\it 9700 AV Groningen} \\[-1.5mm]
  {\normalsize\it THE NETHERLANDS} \\[-1mm]
  {\normalsize\tt aenter@phys.rug.nl} \\[-1mm]
\\ [-1mm]
  {\normalsize Tim Hulshof} \\[-1mm]
  {\normalsize\it Star Numanstraat 93 b} \\[-1.5mm]
  {\normalsize\it 9714 JM Groningen} \\[-1.5mm]
  {\normalsize\it THE NETHERLANDS} \\[-1mm]
  {\normalsize\tt W.J.T.Hulshof@student.rug.nl} \\[-1mm]
{\protect\makebox[5in]{\quad}}}
\begin{document}
\maketitle \baselineskip=14pt \noindent {\bf Abstract.}
In this note we analyse an anisotropic, two-dimensional
 bootstrap percolation model introduced by Gravner and Griffeath.
We present upper and lower bounds
on the finite-size effects. We discuss the similarities with
the semi-oriented model introduced by Duarte.

\bigskip

\newtheorem{theorem}{Theorem} 
\newtheorem{lemma}{Lemma} 
\newtheorem{proposition}[lemma]{Proposition}
\newtheorem{corollary}[lemma]{Corollary}
\newtheorem{definition}[theorem]{Definition}
\newtheorem{conjecture}[theorem]{Conjecture}
\newtheorem{claim}[theorem]{Claim}
\newtheorem{observation}[theorem]{Observation}
\def\proof{\par\noindent{\it Proof.\ }}
\def\reff#1{(\ref{#1})}

\let\zed=\bbbz 
\let\szed=\bbbz 
\let\IR=\bbbr 
\let\R=\bbbr 
\let\sIR=\bbbr 
\let\IN=\bbbn 
\let\IC=\bbbc 

\def\nl{\medskip\par\noindent}

\def\scrb{{\cal B}}
\def\scrg{{\cal G}}
\def\scrf{{\cal F}}
\def\scrl{{\cal L}}
\def\scrr{{\cal R}}
\def\scrt{{\cal T}}
\def\pfin{{\cal S}}
\def\prob{M_{+1}}
\def\cql{C_{\rm ql}}
\def\bydef{\stackrel{\rm def}{=}} 
\def\qed{\hbox{\hskip 1cm\vrule width6pt height7pt depth1pt \hskip1pt}\bigskip}
\def\remark{\medskip\par\noindent{\bf Remark:}}
\def\remarks{\medskip\par\noindent{\bf Remarks:}}
\def\example{\medskip\par\noindent{\bf Example:}}
\def\examples{\medskip\par\noindent{\bf Examples:}}
\def\nonexamples{\medskip\par\noindent{\bf Non-examples:}}

\newenvironment{scarray}{
          \textfont0=\scriptfont0
          \scriptfont0=\scriptscriptfont0
          \textfont1=\scriptfont1
          \scriptfont1=\scriptscriptfont1
          \textfont2=\scriptfont2
          \scriptfont2=\scriptscriptfont2
          \textfont3=\scriptfont3
          \scriptfont3=\scriptscriptfont3
        \renewcommand{\arraystretch}{0.7}
        \begin{array}{c}}{\end{array}}

\def\wspec{w'_{\rm special}}
\def\mup{\widehat\mu^+}
\def\mupm{\widehat\mu^{+|-_\Lambda}}
\def\pip{\widehat\pi^+}
\def\pipm{\widehat\pi^{+|-_\La\bibitem{mi}

mbda}}
\def\ind{{\rm I}}
\def\const{{\rm const}}

\bibliographystyle{plain}


\newpage

\maketitle
\section{Introduction}

Bootstrap percolation (BP) models, sometimes also called k-core percolation or
threshold growth models,
are simple Cellular Automata with
a percolation configuration chosen as a random initial condition.
The deterministic dynamics is given by repeatedly applying a growth rule,
and the question of interest is if the whole system will be occupied in the
long run.
They have been a topic of interest in fields as diverse as
physics -- for the study of magnetic models, metastability, rigidity,
fluid flow in porous media and glassy dynamics --, mathematics,
computer science, neural science and economics,
see e.g. \cite{Ad, AdLe, AL, BB,BFT1,BFT2, CC, CRV, dGDGL1,dGDGL2,
Ho1,Ho2,KWGH,KPBBF,LV,L,RS,SPS,Sch1, TCNG}.

Most of the models which have been considered in the literature are
isotropic. In this paper we consider an anisotropic model in two dimensions
which was introduced by Gravner and Griffeath \cite{GG1,GG2}. Their
model is critical, meaning that there are infinite sets with an infinite
complement which can fill the entire lattice, but no finite set can do so,
and all sets with a finite complement will fill up the lattice.
It has a trivial percolation threshold $p_c= 0$. We determine the
asymptotic behaviour of the finite-size effects, and find it to be
different from the one proposed by Gravner and Griffeath. We discuss the
similarity with the behaviour of the semi-oriented (Duarte) model which is also
anisotropic, and the behaviour of which was established by Mountford.
\cite{Du,ADE,ADE1,Sch2,Mo1,Mo2}.
\smallskip

\smallskip


\section{Definition and some known properties of bootstrap percolation models}

We will consider BP models either on $\bbZ^2$ or on finite subsets thereof.

\smallskip

\noindent
The dynamics is defined as follows:

\smallskip

\noindent
Once a site is occupied, it will remain so forever.

\noindent
For each empty site it is checked at each time step whether in a prescribed
neighborhood $ x+ \mathcal{N}$ of the site $x$ at least $\theta$ sites are
occupied. (The parameter $\theta$ is called the threshold value.)
If so, at the next time step the site will be occupied.

\noindent
At the next time step this procedure is repeated.

\smallskip

Thus this updating rule provides a deterministic, parallel dynamics.
This dynamics runs until no changes occur anywhere anymore.

 To describe the evolution of an occupied
set $A\subset\mathbb{Z}^{2}$ we define
 the discrete time step operator $\mathcal{T}$:
\begin{equation}
\mathcal{T}(A)=A\cup\lbrace x:|(x+\mathcal{N})\cap A|\geq \theta\rbrace.
\end{equation}
Starting from some initial set $A_{0}\subset \mathbb{Z}^{2}$, we
iterate: $\mathcal{T}(A_{n})=A_{n+1}$.
We also define
$A_{\infty}=\mathcal{T}^{\infty}(A_{0})=\bigcup_{n=0}^{\infty}A_{n}$. We
 call a model critical if there exists some infinite $A_{0}$
(but no finite one) with an infinite
complement in $\mathbb{Z}^{2}$ such that
$A_{\infty} = \mathbb{Z}^{2}$, that is
the dynamics eventually fills
all of $\mathbb{Z}^{2}$.

We choose for our random initial configuration a percolation configuration,
where each site is independently occupied with probability $p$.

The standard (isotropic)
model is defined by taking as the neighborhood of a site its four
nearest neighbours and $\theta=2$.

\begin{tabular}{ccccccc}
\, & \, & \, & \, & $\bullet$ & \, & \, \\
$\mathcal{N}$ & $=$ & \ & $\bullet$ & $0$ & $\bullet$ \\
\, & \, & \, & \, & $\bullet$ & \, & \,
\end{tabular}

In that case $p_c=0$ for the
infinite lattice \cite{E} (so for each initial density the lattice will
fill up). The finite-size effects at small $p$ are such that if the
size of a square is of order  $\exp (C_1 \times \frac{1}{p})$,
with $C_1 \gneq  Cst$ for an explicit, computable
constant $Cst$, the square tends to be occupied in the end,
while if it is of order $\exp (C_2 \times \frac{1}{p})$,
with $C_2 \lneq Cst$, it tends to remain mostly empty \cite{AL,Ho1} with
overwhelming probability.

In the Gravner-Griffeath (GG) model, the neighborhood of a site consists
of six sites, to wit,
the closest two sites to the East, the closest
two sites to the West, and the nearest
neighbours in the North and South direction. The threshold value
$\theta$ now is defined to be $3$.

\begin{tabular}{ccccccc}
\, & \, & \, & \, & $\bullet$ & \, & \, \\
$\mathcal{N}$ & $=$ & $\bullet$ & $\bullet$ & $0$ & $\bullet$ & $\bullet$ \\
\, & \, & \, & \, & $\bullet$ & \, & \,
\end{tabular}

 For later reference we also mention the Duarte (semi-oriented) BP
model, in which the neighborhood of a site consists of three sites, namely
its North, West and South neighbours, and $\theta$ is chosen to be $2$.

\begin{tabular}{cccc}
\, & \, & \, & $\bullet$\\
$\mathcal{N}$ & $=$ & $\bullet$ & $0$ \\
\, & \, & \, & $\bullet$
\end{tabular} \newline

This model can not grow in the Western direction.

The GG model (similarly to the Duarte model) has
easy and hard growth directions.
Indeed, once an $N$ by $N$ square
is occupied, a single occupied site at distance $1$ or $2$
will fill the next line segment, in the East or West direction, while
for growth in the North or South direction two sites with no more than 3
empty sites in between need to be occupied, which is much less probable
(locally the probability is of order $p^2$ instead of order $p$).

Two important notions in the theory are those of a ``critical droplet''
(sometimes called nucleation droplet)
and a volume being ``internally spanned''.

A critical droplet is an occupied configuration which will keep on
growing with large probability. In the regular model,
for example, an occupied square
with side length $C_1 \times \frac{1}{p}$, $C_1$ large, has this property,
while in the Duarte model, an occupied line segment of length
$C_1\times \frac{1}{p}\times \ln \frac{1}{p}$, $C_1$ large,
is known to have this property.

The second important notion is that of a volume being internally spanned.
We  will call a volume $V$ `\textit{internally spanned}'
if given some occupied set $A_{0}\subset V$, $A_{\infty}$ occupies
all the sites in $V$.
In that case the initially occupied set $A_{0}$ -- which typically  will be a 
random set -- spans the volume $V$. 
(In the Duarte model the notion of internally spanned is defined 
slightly differently.)

\smallskip
{\bf Remark:} A related quantity, for which a similar scaling behaviour
can be derived, as a corollary of our results,
is the first passage time at the origin, that is the time
at which the origin is occupied for the first time, see e.g.
\cite{AL,GG1}.

\section{Main result and proof}

Our main result is that the volume size for which the Gravner-Griffeath model
changes from typically not being internally spanned to typically
being internally spanned scales as
$\exp ( O( \frac{1}{p}\times {\ln}^{2} (\frac{1}{p})))$.
 
Indeed we have:

\begin{theorem}

Consider the Gravner-Griffeath model.
\newline 
A) If a square is larger than
$\exp ( C_1( \frac{1}{p}\times {\ln}^{2} (\frac{1}{p})))$, with $C_1$ large enough,
it is internally spanned with large probability, for $p$ small enough.
\newline \newline
B) If a square is smaller than
$\exp ( C_2( \frac{1}{p}\times {\ln}^{2} (\frac{1}{p})))$, with $C_2$ small enough,
it is not internally spanned with large probability, for $p$ small enough.

\end{theorem}

{\bf Proof:}
We will prove A) and B) separately.

\smallskip

 To prove A), we will first show that an occupied rectangle of size
\noindent
$2$ by $C_1 \times \frac{1}{p} \times \ln \frac{1}{p}$ is a critical droplet.

Indeed, let such rectangle (a double segment,
long in the North-South direction, width 2 in the East-West direction)
be occupied, then we will first argue that with large probability it will
grow in the Eastern and Western directions, until it fills up a rectangle
of size
\noindent
$2 {(\frac{1}{p})}^{C}$ by $C_1 \times \frac{1}{p} \times \ln \frac{1}{p}$
\noindent
as long as $C$ is smaller than $C_1$. (When $C$ is larger than 1,
such a rectangle is much larger in the East-West than the
North-South direction).

For this to happen, it is sufficient that in all the
$(\frac{1}{p})^C$ line segments on the
left and on the right at least one site is occupied in the
original configuration. But this happens with probability
${(1-{(1-p)}^{C_1 \frac{1}{p} \ln \frac{1}{p}})}^{O({\frac{1}{p}}^{C})} 
\simeq {(1-p^{C_{1}})}^{O( {\frac{1}{p}}^{C})}$,
which is close to one when $C$ is smaller than $C_1$.

\smallskip

Once the larger rectangle is occupied, with $C$  chosen to be large enough
(it should be at least larger than $2$), it will grow in all 
directions, with large probability.
Indeed, if we take the rectangle to be $N_1$ by $N_2$, $N_1$ much larger than
$\frac{1}{p}$ and $N_2$ much larger than ${(\frac{1}{p})}^{2}$, then
the probability that the rectangle does not stop growing in either 
horizontal or vertical direction is larger than

\noindent
$\prod_{k=1}^{\infty} (1-(1-p^{2})^{N_2+k })(1-(1-p)^{N_{1}+k})$.

Minus the logarithm of this probability is approximately

\noindent
$\sum_{k=1}^{\infty} (1-p^{2})^{N_2+k}+ (1-p)^{N_{1}+k}$ which is small,
showing that the corresponding probability is close to $1$.

\smallskip

As the probability for a particular rectangle to be a critical droplet 
(thus the density of critical droplets) is larger than
$p^{2 C_1 \frac{1}{p} \ln \frac{1}{p}}=
\exp (- 2 C_1 \frac{1}{p} {\ln}^{2} (\frac{1}{p}))$, the minimal 
size necessary for a square to contain with large probability a 
critical droplet is bounded from
above by the inverse of this density. This critical droplet then will grow
to fill up the entire square. Indeed, wherever in the square the droplet is, 
for small $p$ there will always be enough space to grow sufficiently far  
in at least two perpendicular directions (remember that 
the size of the square is much larger than $\frac{1}{p^{2}}$) and 
once the occupied set is big enough, with large probability it will then 
fill up its complement in the square.

\smallskip

\noindent

This finishes the proof of part A).

\smallskip

To prove part B), we essentially follow the analysis of Gravner and Griffeath
\cite{GG1}, and point out where it needs to be modified.
(Gravner \& Griffeath based this part of their analysis
largely on the proof of the similar statement for ordinary bootstrap
percolation as given by Aizenman \& Lebowitz in \cite{AL}.)

\smallskip

\noindent
We will first quote their definition and main claim:

{\bf Definition:}
[We] call a rectangle $R$ in $\mathbb{Z}^2$ \textit{potentially
internally spanned} (PIS) if it is either a single site in $\Pi (p)$ or
if (\textit{i}) for every (integer) vertical $\ell$ through $R$
there exist two sites $x,y \in \Pi (p) \cap R$ such that
$\lVert x-y \rVert_{\infty} \leq 4$ and they are
both at $\ell^{\infty}$-distance at most 4
from $\ell$, and (\textit{ii}) every horizontal
line $\ell$ through it has a site in $\Pi(p) \cap R$
at $\ell^{\infty}$-distance at most 2 from $\ell$.

\smallskip

\noindent
Here $\Pi(p)$ denotes the set of initially occupied sites, when
the occupation probability is $p$. The property PIS is obviously weaker
than that of being internally spanned.

\smallskip

\textbf{Claim:} Let $L < M$ be positive integers. Assume
that the origin is not eventually occupied if the dynamics are
restricted to $[-L,L]^{2}$, but is eventually occupied if
the dynamics are restricted to $[-M,M]^{2}$. For every integer
$a \in [4, L/4]$, there exists a PIS rectangle $R$ included
in $[-M,M]^{2}$ whose longest side is between $a$ and $4a$.
The proof to the claim is given in \cite{GG1}.
\\[3mm]
From the definition of PIS it is then derived that the following must hold:

\begin{align}
P(k,l)
&
:= P(\text{a fixed}\: k \times l\: \text{rectangle is PIS})\nonumber\\
&
\leq (1-(1-p^2)^k)^l (1-(1-p)^l)^k \nonumber\\
&
\leq \min\lbrace(1-(1-p^2)^k)^l,(1-(1-p)^l)^k \rbrace \nonumber \\
&
:=\min\lbrace I,II \rbrace \nonumber \\
\end{align}

Now we deviate from \cite{GG1}. We choose
$k= {\frac{1}{p^{3/2}}}$, and $l = C_{2} \frac{1}{p} \ \ln \frac{1}{p}$,
then we have

\begin{align}
II
&
= (1-(1-p)^{\frac{C_{2}}{p} \ln \frac{1}{p}} )^{\frac{1}{p^{3/2}}}
\nonumber \\
&
\simeq (1-p^{C_{2}})^{\frac{1}{p^{3/2}}} \nonumber\\
&
\simeq \exp(-\frac{p^{C_{2}}}{p^{3/2}}) \nonumber\\
&
\leq \exp(-C_{2} \frac{1}{p} \ln^{2} \frac{1}{p})
\end{align}
 
if $C_2$ is small enough, and moreover

\begin{align}
I
&
= (1-(1-p^2)^{\frac{1}{p^{3/2}}})^{C_{2} \frac{1}{p} \ln \frac{1}{p}} \nonumber \\
&
\simeq (1-e^{-\sqrt{p}})^{C_{2} \frac{1}{p} \ln\frac{1}{p}} \nonumber \\
&
\simeq p^{\frac{C_{2}}{2p} \ln \frac{1}{p}} \nonumber \\
&
= \exp(-\frac{C_{2}}{2p} \ln^{2} \frac{1}{p})
\end{align}

which shows that for this choice of $k$ and $l$ and $C_{2}$ small enough

\begin{equation}
P(k,l) \leq \exp(-\frac{C_{2}}{2p} \ln^{2} \frac{1}{p}).
\end{equation}

This inequality contradicts the claim on the asymptotics
of Gravner and Griffeath.
(We suspect that in their derivation, which they don't give completely but
describe as a ``straightfoward computation'', the following went wrong:
They seem to have used the approximation ${(1-x)}^{N} = \exp (-xN)$ not
only for small values of $x$, but also for larger $x$-values.
For example, in an expression like $I$,
with the choice $x ={(1-p^2)}^{\frac{1}{p^{3/2}}}$, and
$N = C_{2} \times \frac{1}{p} \times \ln \frac{1}{p}$, 
the approximation gives a wrong
answer, which is of the form proposed in \cite{GG1}.
For some further discussion on this point, and another derivation of
part A) of the theorem, see \cite{Hul}.)

Thus if the size of a square is less than the inverse of this probability,
it will not be PIS with large probability. This is the main ingredient
to prove B). For further details see again \cite{GG1}.

\section{Discussion and conclusions}

We have found that the asymptotics of the finite-size effects is different
from the one proposed in \cite{GG1}.
In fact, the behaviour of the GG model is quite similar
to that of the Duarte model.

It has the same asymptotics, and whereas the Duarte model has a single line
segment of length $ C_1 \times \frac{1}{p}\times \ln \frac{1}{p}$,
with $C_1$ large enough, as its critical droplet (first observed by Schonmann
and communicated to the authors of \cite{ADE,ADE1}),
in the GG model a double segment of the same length plays the same role.
Therefore the difference between the two models suggested in \cite{GG1}
(one model understandable via its critical droplets, the other one via
its growth mechanism) is in our opinion spurious. This observation leads us to
expect that in considerable generality
critical anisotropic BP models with slow and fast
directions will have the same asymptotics for their finite-size effects.

{\bf Open questions:} \newline
It would be interesting to see if it is possible to identify the exact
asymptotic constant (in other words, can $C_1$ and $C_2$ be chosen to
be identical, as happens in the isotropic model \cite{Ho1}?).

Also it seems a natural question
to see what the behaviour of anisotropic models
in higher dimensions might be.

\smallskip

However, at present we have no results pertaining to these questions.

\bigskip

{\bf Acknowledgements:} We thank Christof K\"ulske for a critical 
reading of the manuscript.

\end{document}